\documentclass[aps,prb,twocolumn,groupedaddress,showpacs]{revtex4}

\usepackage{graphicx}% Include figure files
\usepackage{amsmath}

\begin{document}

%\preprint{}

\title{Josephson Plasma in RuSr$_{2}$GdCu$_{2}$O$_{8}$}

% Please use the appropriate macro for each type of information

\author{H. Shibata}
\email[]{shibata@will.brl.ntt.co.jp}
%\thanks{}
\affiliation{NTT Corporation, NTT Basic Research Laboratories, 3-1 Morinosato Wakamiya,
Atsugi-shi, Kanagawa 243-0198, Japan}


\date{\today}

\begin{abstract}
Josephson plasma in RuSr$_{2}$GdCu$_{2}$O$_{8}$, Ru$_{1-x}$Sr$_{2}$GdCu$_{2+x}$O$_{8}$ (x = 0.3), 
and RuSr$_{2}$Eu$_{2-x}$Ce$_{x}$Cu$_{2}$O$_{10}$ (x = 0.5) compounds is investigated by the sphere 
resonance method.
The Josephson plasma is observed in a low-frequency region 
(around 8.5 cm$^{-1}$ at T $\ll$ $T_{c}$ ) for ferromagnetic 
RuSr$_{2}$GdCu$_{2}$O$_{8}$, while it increases to 35 cm$^{-1}$ 
for non-ferromagnetic Ru$_{1-x}$Sr$_{2}$GdCu$_{2+x}$O$_{8}$ (x = 0.3), 
which represents a large reduction in the Josephson coupling at 
ferromagnetic RuO$_{2}$ block layers.
The temperature dependence of the plasma does not shift 
to zero frequency ({\it i.e.} $j_{c}$ = 0) at low temperatures, 
indicating that there is no transition from the $0$-phase to the $\pi$-phase in these compounds.
The temperature dependence and the oscillator strength of the peak are different from those of 
other non-magnetic cuprates, and the origins of these anomalies are discussed.
\end{abstract}

\pacs{74.25.Gz, 74.72.Jt, 74.80.Dm}

%\keywords{}

\maketitle

%\section{}
%\subsection{}
%\subsubsection{}

It is believed that the electronic state of high-$T_{c}$ cuprates 
in the superconducting state is two-dimensional. 
Many cuprates exhibit dc and ac intrinsic Josephson effects, 
which reveals that the $\cdots$S/I/S/I/S/I/S$\cdots$ 
($\cdots$superconductor/insulator/superconductor$\cdots$)
-type Josephson-coupled layer model is applicable to high-$T_{c}$ cuprates.
In this case, the plasma of condensed carriers observed in most 
of the cuprates in the far infrared to microwave region is thought to be 
Josephson plasma.
There have been many studies of this phenomenon, since it represents the interlayer phase coherence 
between CuO$_{2}$ layers and may be related to the mechanism of high-$T_{c}$ superconductivity
\cite{matsuda,shibsphere,shiblconf26,anderson,shibDJPR2,uchidaDJPR,marelDJPR}.
Of particular importance is that it provides information about the junction parameters 
without the need to fabricate the mesa of an intrinsic Josephson junction; 
the maximum critical current $j_{c}$ of the junctions is directly related to the Josephson 
plasma frequency $\omega_{p}$ by the equation 
$\omega_{p}^{2}$ = 8$\pi^{2}$$cd$$j_{c}$/$\epsilon_{0}$$\Phi_{0}$, 
where $d$ and $\epsilon_{0}$ are the width and dielectric constant of the insulating layer.
It should be noted that $\omega_{p}$ can be deduced even from the ceramics, since 
the Josephson plasma can be observed by measuring the sphere resonance of powder samples 
as well as by conventional single crystal measurements\cite{noh,noh2,shibsphere,shibDJPR,shibbi}.

Recently, much attention has been paid to RuSr$_{2}$GdCu$_{2}$O$_{8}$ and 
related materials, as the coexistence of superconductivity and ferromagnetism 
has been reported in these compounds\cite{bauernfeind,bernhard,felner,klamut,boris}.
Since superconductivity and ferromagnetism are mutually exclusive, 
many possible superconducting order parameters have already been discussed, 
including the self-induced vortex state, the Fulde-Ferrell-Larkin-Ovchinnikov type, triplet 
superconductivity, and the $\pi$-phase\cite{bauernfeind,bernhard,felner,klamut,pickett,ohashi,buzdin}.
The crystal structure of RuSr$_{2}$GdCu$_{2}$O$_{8}$ is similar to that of 
YBa$_{2}$Cu$_{3}$O$_{7-\delta}$ with RuO$_{2}$ layers replacing the Cu-O chains, and 
the ferromagnetic order, which is observed at a Curie temperature of $T_{\text M}$ = 133 K, 
is attributed to the Ru moment.
Although the magnetic structure of the compound has not yet been clarified, 
it seems that the Ru moments are anti-ferromagnetic ordered along the c-axis 
and canted to the ab-plane, which makes the ab-plane weak-ferromagnetic\cite{jorgensen,lynn}.
So, the system may be regarded as the $\cdots$S/F/S/F/S/F/S$\cdots$ 
($\cdots$superconductor/(ferro)magnet/superconductor$\cdots$)  
-type Josephson-coupled multilayers along the c-axis. 
In this multilayer system, the $\pi$-phase, which has a superconducting order parameter that changes 
the phase by $\pi$ between two adjacent superconducting layers, 
seems to be realized since the node at the 
ferromagnetic layer greatly reduces the pair breaking effects.
The model calculations for this compound predict a transition from the 0-phase to the $\pi$-phase 
at low temperatures\cite{pickett,ohashi,buzdin}, and a peculiar temperature dependence of 
$j_{c}$; as the temperature decreases, $j_{c}$ should achieve its maximum value, decrease to zero 
at the transition line due to the decoupling of the junctions, and then increase again in the 
$\pi$-phase region\cite{buzdin}.
It should noted that this $j_{c}$ temperature dependence has been realized experimentally 
in artificial Josephson junctions consisting of Nb and ferromagnetic Cu$_{x}$Ni$_{1-x}$ alloy 
as the transition from the 0-junction to the $\pi$-junction\cite{ryazanov}.
In the present case, it is impossible to determine $j_{c}$ by I-V measurement, 
since no one has yet grown millimeter size single crystal of Ru-cuprates\cite{lin}.
The only way to determine the $j_{c}$ temperature dependence seems to be to measure $\omega_{p}$ by measuring 
the sphere resonance of powder samples.

This paper reports our measurement of the far-infrared sphere resonance of RuSr$_{2}$GdCu$_{2}$O$_{8}$, 
Ru$_{1-x}$Sr$_{2}$GdCu$_{2+x}$O$_{8}$ (x = 0.3), and RuSr$_{2}$Eu$_{2-x}$Ce$_{x}$Cu$_{2}$O$_{10}$
 (x = 0.5) powder samples down to 7 cm$^{-1}$.
While all the samples exhibit the Josephson plasma resonance, the temperature dependence does not shift 
to zero frequency at low temperature as expected from the $\pi$ transition. 
The peak frequency is greatly reduced in the ferromagnetic samples, indicating a considerable reduction 
in the Josephson coupling of the junctions.
The anomalous features of the plasma are also discussed.

% format so that long equations can be displayed. Use
%\begin{widetext}
%\end{widetext}

% figures should be put into the text as floats.
% Use the graphics or graphicx packages (distributed with LaTeX2e)
% and the \includegraphics macro defined in those packages.
% See the LaTeX Graphics Companion by Michel Goosens, Sebastian Rahtz,
% entire page. There is no need to do explicit centering.

\begin{figure}
\includegraphics[width=216pt]{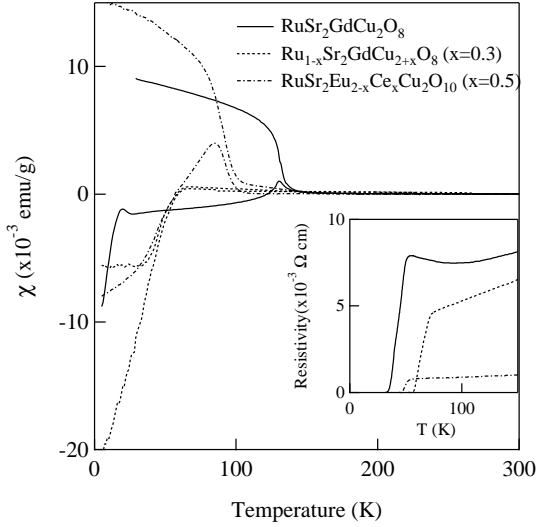}%
\caption{\label{mpms} Magnetic susceptibility and resistivity (inset) of 
RuSr$_{2}$GdCu$_{2}$O$_{8}$, Ru$_{1-x}$Sr$_{2}$GdCu$_{2+x}$O$_{8}$ (x = 0.3) 
and RuSr$_{2}$Eu$_{2-x}$Ce$_{x}$Cu$_{2}$O$_{10}$ (x = 0.5) ceramics.}
\end{figure}

The ceramic samples were synthesized by the conventional solid-state reaction of oxides and 
carbonates under almost the same conditions described in previous reports
\cite{bauernfeind,bernhard,felner,klamut,bernhard2}.
After several cycles of sintering and regrinding, annealing was performed at 1065$^{\circ}$C 
for 150 hr at 1 atm O$_{2}$ for RuSr$_{2}$GdCu$_{2}$O$_{8}$, at 600$^{\circ}$C for 50 hr at 400 atm 
O$_{2}$ for Ru$_{1-x}$Sr$_{2}$GdCu$_{2+x}$O$_{8}$ (x = 0.3), and at 600$^{\circ}$C for 30 hr at 180 
atm O$_{2}$ for RuSr$_{2}$Eu$_{2-x}$Ce$_{x}$Cu$_{2}$O$_{10}$ (x = 0.5).
A furnace for hot isostatic pressing (HIP) was used for the high-oxygen-pressure annealing.
The magnetic susceptibility and resistivity of the samples are summarized in Fig.~\ref{mpms}. 
Although a magnetic transition is observed at $T_{\text M}$ = 133 K for RuSr$_{2}$GdCu$_{2}$O$_{8}$ 
and around 100 K for RuSr$_{2}$Eu$_{2-x}$Ce$_{x}$Cu$_{2}$O$_{10}$ (x = 0.5), 
no magnetic transition is observed for Ru$_{1-x}$Sr$_{2}$GdCu$_{2+x}$O$_{8}$ (x = 0.3).
The resistivity shows $T_{c}^{\text{onset}}$ = 53 K and $T_{c}^{\text{zero}}$ = 34 K for 
RuSr$_{2}$GdCu$_{2}$O$_{8}$,  $T_{c}^{\text{onset}}$ = 64 K and  $T_{c}^{\text{zero}}$ = 56 K 
for Ru$_{1-x}$Sr$_{2}$GdCu$_{2+x}$O$_{8}$ (x = 0.3), and  $T_{c}^{\text{onset}}$ = 54 K and  
$T_{c}^{\text{zero}}$ = 46 K for RuSr$_{2}$Eu$_{2-x}$Ce$_{x}$Cu$_{2}$O$_{10}$ (x = 0.5). 
The powder X-ray diffraction indicated a single phase for 
RuSr$_{2}$GdCu$_{2}$O$_{8}$, while a slight trace of an unknown impurity phase (less than 5 \%) was 
observed for Ru$_{1-x}$Sr$_{2}$GdCu$_{2+x}$O$_{8}$ (x = 0.3) and 
RuSr$_{2}$Eu$_{2-x}$Ce$_{x}$Cu$_{2}$O$_{10}$ (x = 0.5).
The magnetic susceptibility, $T_{c}$, magnitude of resistivity, and x-ray results for 
these data agree well with previous reports\cite{bauernfeind,bernhard,felner,klamut,bernhard2}.
The samples were ground into fine particles about 2 $\mu$m in diameter, mixed with 
polyethylene powder, and then pressed into pellets about 2 mm thick. 
Transmission spectra of the pellets were measured down to 7 cm$^{-1}$ using a Fourier 
transform interferometer combined with a Si bolometer.

% Surround figure environment with turnpage environment for landscape
% figure
% \begin{figure}
% \end{figure}

\begin{figure}
\includegraphics[width=216pt]{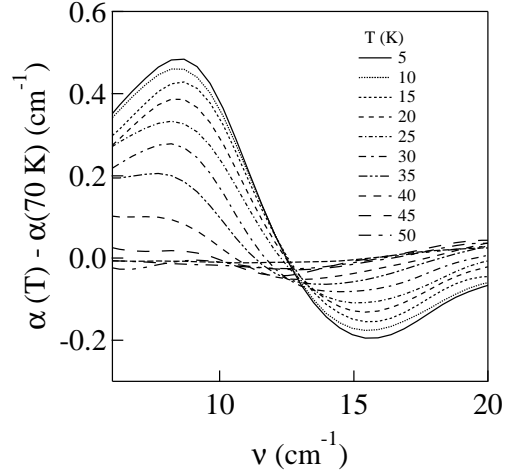}%
\caption{\label{ru} Difference between $\alpha$ in the superconducting and normal states for 
RuSr$_{2}$GdCu$_{2}$O$_{8}$ ceramics.}
\end{figure}

Figure~\ref{ru} shows the difference between the absorption coefficients of the superconducting and 
normal states for RuSr$_{2}$GdCu$_{2}$O$_{8}$ ceramics.
Below $T_{c}^{\text{onset}}$, the Josephson plasma peak appears 
and the oscillator strength increases as the temperature decreases.
The peak is very broad and it is impossible to determine the peak frequency, 
however it becomes rather narrow below $T_{c}^{\text{zero}}$ = 34 K. 
The peak is around 8.5 cm$^{-1}$ at 5 K. 
The peak frequency is very low compared to that of YBa$_{2}$Cu$_{3}$O$_{7-\delta}$ plasma 
above 100 cm$^{-1}$, which has a similar crystal structure and is in a similar doping level 
(optimum to overdoped region)\cite{shibsphere,tokunaga,kuma,liu}, and this suggests a large reduction 
in the Josephson coupling at the ferromagnetic RuO$_{2}$ block layers.
The peak does not shift to zero frequency with decreasing temperature, which indicates that 
there is no 0 - $\pi$ transition in this compound.

\begin{figure}
\includegraphics[width=216pt]{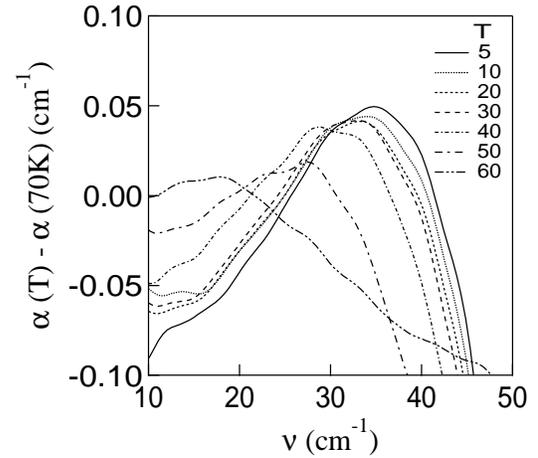}%
\caption{\label{rux} Difference between $\alpha$ in the superconducting and normal states for 
Ru$_{1-x}$Sr$_{2}$GdCu$_{2+x}$O$_{8}$ (x = 0.3) ceramics.}
\end{figure}

Figure~\ref{rux} shows the Josephson plasma peaks of non-ferromagnetic
Ru$_{1-x}$Sr$_{2}$GdCu$_{2+x}$O$_{8}$ (x = 0.3) ceramics.
Compared with ferromagnetic RuSr$_{2}$GdCu$_{2}$O$_{8}$, 
the peak frequency increases greatly to 35 cm$^{-1}$ at 5 K. 
This clearly indicates the large reduction in the Josephson coupling at the ferromagnetic
RuO$_{2}$ block layers for RuSr$_{2}$GdCu$_{2}$O$_{8}$ and its absence at the non-ferromagnetic 
(Ru, Cu)O$_{2}$ block layers for Ru$_{1-x}$Sr$_{2}$GdCu$_{2+x}$O$_{8}$ (x = 0.3).
The peak frequencies increase as the temperature decreases, 
indicating the absence of a 0 - $\pi$ transition in this compound.
The peak oscillator strength is very weak compared with other cuprates, as discussed later.

\begin{figure}
\includegraphics[width=216pt]{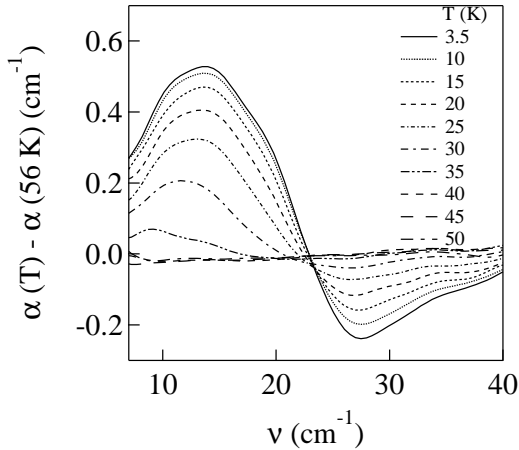}%
\caption{\label{ru1222} Difference between $\alpha$ in the superconducting and normal states for 
RuSr$_{2}$Eu$_{2-x}$Ce$_{x}$Cu$_{2}$O$_{10}$ (x = 0.5) ceramics.}
\end{figure}

A similar peak is also observed in RuSr$_{2}$Eu$_{2-x}$Ce$_{x}$Cu$_{2}$O$_{10}$ (x = 0.5) 
ceramics, as shown in Fig.~\ref{ru1222}. 
In this case, the peak frequency is around 13 cm$^{-1}$ at 5 K, corresponding to the 
large reduction in the coupling at the magnetic RuO$_{2}$ block layers.
There are two kinds of junction per unit cell in the compound; 
that at the RuO$_{2}$ block layers and that at the fluorite-type (Eu, Ce)O$_{2}$ block layers, 
and we may expect the existence of double Josephson plasma\cite{shibDJPR,shibDJPR2,uchidaDJPR,marelDJPR}, 
which is not seen in Fig~\ref{ru1222}.
The absence of double Josephson plasma may be explained by the degenerating plasma frequencies 
of both junctions, since the plasma frequency of fluorite-type block layers is around 
10 cm$^{-1}.$\cite{shiblconf26}
The observed broadness of the peak also supports this explanation.
Moreover, the temperature dependence of the peak does not show the 0 - $\pi$ transition.

Although there is no transition to the $\pi$-phase, 
the plasma shows certain anomalies that are quite different from those of other non-magnetic cuprates.
The first is the temperature dependence of the plasma in RuSr$_{2}$GdCu$_{2}$O$_{8}$; 
the peak frequency of the plasma is almost constant 
and only the peak oscillator strength increases as the temperature decreases 
below $T_{c}^{\text{zero}}$, which is in contrast to the monotonic $\omega_{p}$ increase
in other cuprates.
Since the peak oscillator strength is proportional to $\omega_{p}^{2}f$, where
$f$ is the volume fraction of the superconductor in the pellet\cite{shibspherec}, 
this indicates that $\omega_{p}$ is almost constant while $f$ increases as the temperature decreases.
The calculation of $\cdots$S/F/S/F/S$\cdots$-type multilayers predicts that $\omega_{p}$ decreases slightly 
at low temperatures from the monotonic increase as the exchange energy increases from zero, and the decrease 
becomes large as the exchange energy increases to the 0 - $\pi$ transition line\cite{buzdin}.
The constant behavior observed for $\omega_{p}$ suggests that the exchange energy, while 
not zero for the sample, is small and far from the 0 - $\pi$ transition line.

The second anomaly is the weak oscillator strength of the plasma; 
the observed oscillator strength of the peak is smaller than that of other non-magnetic cuprates, 
especially for Ru$_{1-x}$Sr$_{2}$GdCu$_{2+x}$O$_{8}$ (x = 0.3), 
while we make the volume fraction $f$ around 0.01 for all pellets.
This can be explained by the very small grain size of the Ru-cuprates,
which is known as granularity\cite{bernhard}.
With the sphere resonance method, the powder (about 2 $\mu$m) in the pellets is assumed to be a single crystal, 
and powder whose c-axis is parallel to the electric field of light causes the peak. 
This assumption seems not to hold for Ru-cuprates since most of the powder is formed of granules rather than a single crystal.
The granular powder, which is dominated by the in-plane dielectric constant, does not form a sphere resonance peak, 
while the powder, which is almost composed of single domains, forms a broad peak due to the modulation of the Josephson-coupling 
from the in-plane dielectric constant.
So, we expect the total absorption peak of granular samples to be small and broad.

The last anomaly is the strong peak broadening between $T_{c}^{\text{zero}}$ and $T_{c}^{\text{onset}}$ 
in RuSr$_{2}$GdCu$_{2}$O$_{8}$.
One possible explanation is the granularity of the samples as discussed above.
However in this case, we cannot explain the broadening change at $T_{c}^{\text{zero}}$.
Another possible explanation is the melting of the self-induced vortex at $T_{c}^{\text{zero}}$, 
which has been suggested by other experiments\cite{bernhard,tokunaga}.
In this scenario, the peak is sharp in a solid phase below $T_{c}^{\text{zero}}$, 
and it becomes broad in a vortex liquid phase between $T_{c}^{\text{zero}}$ and $T_{c}^{\text{onset}}$.
Here, we also expect the peak to become sharp again in the liquid phase, 
which is different from the experiment.

% tables should appear as floats within the text
% The ruledtabular enviroment adds doubled rules to table and sets a
% reasonable default table settings.
% Use the table* environment to get a full-width table in two-column
% Add \usepackage{longtable} and the longtable (or longtable*}
% environment for nicely formatted long tables. Or use the the [H]
% placement option to break a long table (with less control than 
% \begin{table}%[H] add [H] placement to break table across pages
% \begin{ruledtabular}
% Lines of table here ending with \\
% \end{tabular}
% \end{ruledtabular}
% \end{table}

% Surround table environment with turnpage environment for landscape
% table
% \begin{table}
% \begin{ruledtabular}
% \end{tabular}
% \end{ruledtabular}
% \end{table}

%\appendix
%\section{}

%While all these samples do not show the 0 - $\pi$ transition as expected, 

In summary, Josephson plasma is observed at 8.5 cm$^{-1}$ and 13 cm$^{-1}$ for ferromagnetic 
RuSr$_{2}$GdCu$_{2}$O$_{8}$ and RuSr$_{2}$Eu$_{2-x}$Ce$_{x}$Cu$_{2}$O$_{10}$ (x = 0.5), 
and it increases to 35 cm$^{-1}$ for non-ferromagnetic 
Ru$_{1-x}$Sr$_{2}$GdCu$_{2+x}$O$_{8}$ (x = 0.3), 
which indicates a large reduction in the Josephson coupling at the 
ferromagnetic RuO$_{2}$ block layers.
Although the temperature dependence of the plasma in these compounds does not show the 0 - $\pi$
phase transition, no increase in $\omega_{p}$ is observed for RuSr$_{2}$GdCu$_{2}$O$_{8}$ 
as the temperature decreases, which suggests a small exchange energy in the compound.
The anomalous temperature dependence of the peak frequency, 
broadening of the peak above $T_{c}^{\text{zero}}$, and the small oscillator strength were also discussed.

%\begin{acknowledgments}
The author thanks Y. Ohashi, Y. Matsuda, M. B. Gaifullin, H. Takayanagi, M. Naito, and A. Matsuda 
for valuable discussions.
%\end{acknowledgments}

\bibliography{refru.bib,refshib.bib}

\end{document}